\documentclass[aps,prl,twocolumn,showpacs,superscriptaddress,groupedaddress]{revtex4}  
\usepackage{graphicx}  
\usepackage{dcolumn}   
\usepackage{bm}        
\usepackage{amssymb}   
\usepackage{amsmath}

\begin{document}
\title{The Virial Theorem in Graphene and other Dirac Materials}
\author{J.~Dustan Stokes} \affiliation{Department of Physics, Boston College, Chestnut Hill, Massachusetts 02467, USA} \affiliation{Brookhaven National Laboratory, Upton, New York 11973, USA}
\author{Hari P.~Dahal} \affiliation{Theoretical Division, Los Alamos National Laboratory, Los Alamos, New Mexico 87545, USA} \affiliation{American Physical Society, 1 Research Road, Ridge, NY 11961, USA}
\author{Alexander V.~Balatsky} \affiliation{Theoretical Division and Center for Integrated Nanotechnology, Los Alamos National Laboratory, Los Alamos, New Mexico 87545, USA}
\author{Kevin S.~Bedell} \affiliation{Department of Physics, Boston College, Chestnut Hill, Massachusetts 02467, USA}
\date{\today}

\begin{abstract}
The virial theorem is applied to graphene and other Dirac Materials for systems close to the Dirac points where the dispersion relation is linear.
From this, we find the exact form for the total energy given by $E = \mathcal{B}/r_s$ where $r_s a_0$ is the mean radius of the $d$-dimensional sphere containing one particle, with $a_0$ the Bohr radius, and $\mathcal{B}$ is a constant independent of $r_s$.
This result implies that, given a linear dispersion and a Coulombic interaction, there is no Wigner crystalization and that calculating $\mathcal{B}$ or measuring at any value
of $r_s$ determines the energy and compressibility for all $r_s$.  In addition to the total energy we calculate the exact forms of the chemical potential, pressure and inverse compressibility in arbitrary dimension.
\end{abstract}

\maketitle

The linear dispersion of graphene near the Dirac points~\cite{Wallace} combined with the experimental realization of
graphene sheets~\cite{Novo} has provided the impetus for countless theoretical and experimental investigations~\cite{AIP}.
Studies of graphene have been fueled by the material's broad spectrum of interesting features---from the pragmatic possible technological implementations
to its relationship to massless fermions travelling at the speed of light, as the
linear dispersion is identical in form to the Dirac Hamiltonian for massless relativistic fermions.
From the point of fundamental physics, one of the most important questions is about the role of electron-electron 
interaction in the system with linear dispersion. This question is more complicated than it sounds for the Dirac Materials, materials that have linearly dispersive 
excitations~\cite{DiracMaterials}, because at least close to the Dirac point, where the Fermi-surface reduces to a point, the density of state vanishes giving rise to a
suppressed Coulomb interaction screening. For reduced density, the Coulomb interaction in conventional electron systems would lead to
the localization of the charge in space, also known as Wigner crystallization. Does this conventional physics hold in linearly dispersive system at the dilute limit?
or does the difference in the dispersion relation introduces unconventional physics?
This question is very relevant because the localization of carriers in graphene has not yet been observed.

Central to the determination of the role of the Coulomb interaction is the measurement and calculation of the electronic compressibility which 
depends upon the total of the potential and kinetic energy. 
The compressibility of monolayer graphene has been measured and the data seems to be well described by the kinetic energy's
contribution alone~\cite{Martin} with effective Fermi velocity which incorporates the effect of the Coulomb interaction. Random Phase Approximation~\cite{Barlas} and Hartree-Fock~\cite{Hwang} studies have concluded
that contributions to the compressibility from the Coulomb interaction yield a (logarithmically) doping dependent correction
of between 10 and 20 percent in the experimental regime. We see significantly different phenomenon viewing the same problem through the lens
of the virial theorem.

Using the virial theorem for the linearly dispersive system, we find that an exact form of the total energy valid for all doping densities can be expressed as
\begin{equation}
\label{soln}
\bar{E}(r_s) = \frac{\mathcal{B}}{r_s}.
\end{equation}
$\mathcal{B}$ is a constant which does not depend on $r_s$, however, the exact numerical value will depend on the material and the dimensionality of the system.  Eq.~(\ref{soln}) is the main result of this work.  The virial also gives concise confirmation of the absence of Wigner crystalization at all densities originally argued in~\cite{HariWigner} for low densities,  and a qualitatively different description of the compressibility than is currently reported in the literature.

The approach we use to obtain Eq.~(\ref{soln}) follows the argument of Argyres~\cite{Argyres}, applied earlier to a quadratic dispersion, with slight but substantive modifications to account for the linear dispersion.  We consider the virial theorem for
a system with a single particle energy that has a linear dispersion relation $E_p =\pm \mathbf{v}\cdot \mathbf{p}$, with $\mathbf{v} = v_g \hat{\mathbf{p}}$ (where in graphene the characteristic velocity $v_g \sim 10^6$ m/s), and interacts with the other electrons and a uniform background charge via
a Coulombic potential $V \propto 1/r$.  It is in this sense that this result is valid for other (Coulombic) Dirac materials found in arbitrary dimension.  The virial has been studied extensively
for the conventional electron gas~\cite{Argyres,March} but has not yet been studied in graphene or other Dirac materials with a linear dispersion.

We focus here on the two dimensional case and introduce the generalized volume $\Omega$, which is the area $A$ in 2D, but the proof applies in arbitrary dimension.  Given a Hamiltonian $H = T + V$, where the kinetic energy is given by $T = \sum_i \mathbf{v}_i\cdot \mathbf{p}_i$, and the potential
energy $V = V_e + V_{eb} + V_{b}$.
$V_e$ is the potential energy between electrons,
\begin{equation}
V_e=\frac{1}{2}\sum_{i\neq j}\frac{e^2}{|\mathbf{r}_i-\mathbf{r}_j|},
\end{equation}
$V_{eb}$ is the potential energy between the electrons and the background of positive charge,
\begin{equation}
V_{eb}=-e\sum_{i}\int_A d^2x\frac{\rho}{|\mathbf{r}_i-\mathbf{x}|},
\end{equation}
and $V_{b}$ is the potential energy of the background of positive charge,
\begin{equation}
V_{b}=\frac{1}{2}\int_A d^2x\int_A d^2x'\frac{\rho^2}{|\mathbf{x}-\mathbf{x}'|}.
\end{equation}
The background positive charge density $\rho \propto r_s^{-2}$ and the area $A \propto r_s^2$.
$V_{e}$ is then a homogeneous function of degree -1 in $\{\mathbf{r}_i\}$,
$V_{b}$ is a homogeneous function of degree -1 in $r_s$, and
$V_{eb}$ is a homogeneous function of degree -1 in $\{\mathbf{r}_i, r_s\}$.
The Euler theorem for homogeneous functions combined with the definition of V yields
\begin{equation}
\label{Vhomo}
\sum_i \mathbf{r}_i \cdot \frac{\partial V}{\partial \mathbf{r}_i}=-V-r_s\frac{\partial V}{\partial r_s}.
\end{equation}
Eq.~\ref{Vhomo} is exactly the same result one gets when calculating the virial for the 2DEG.

The difference lies in the kinetic term, for the linear (quadratic) dispersion in graphene (2DEG) the kinetic energy is a homogeneous function
of degree 1 (2).  Thus, for graphene
\begin{equation}
\label{Thomo}
\sum_i \mathbf{p}_i \cdot \frac{\partial T}{\partial\mathbf{p}_i} = T.
\end{equation}

We consider the virial
\begin{equation}
G=\sum_i\mathbf{r}_i\cdot \mathbf{p}_i
\end{equation}
the quantum mechanical time derivative of which is
\begin{eqnarray}
\dot{G}&=&\frac{1}{i\hbar}[G,H] \nonumber \\
&=&\frac{1}{i\hbar}\left(\sum_i[\mathbf{r}_i,T]\cdot \mathbf{p}_i + \sum_i \mathbf{r}_i\cdot [\mathbf{p}_i,V]\right) \nonumber \\
&=&\sum_i \mathbf{p}_i \cdot \frac{\partial T}{\partial \mathbf{p}_i} - \sum_i \mathbf{r}_i \cdot \frac{\partial V}{\partial \mathbf{r}_i} \nonumber \\
&=&T + V + r_s \frac{\partial V}{\partial r_s}. \label{VirialNoExp}
\end{eqnarray}
The last equality follows from Eqs.~(\ref{Vhomo}) and~(\ref{Thomo}).
Because T is independent of $r_s$, $r_s\frac{\partial V}{\partial r_s}=r_s \frac{\partial H}{\partial r_s}$.
We then take the expectation value of Eq.~(\ref{VirialNoExp}), note that $\overline{[G,H]}=0$ and find~\footnote{This is for free, periodic and open boundary conditions, as proven in~\cite{Argyres}.}
\begin{equation}
\label{lemotjuste}
\bar{T} + \bar{V} = -r_s \frac{\partial \bar{E}}{\partial r_s}.
\end{equation}
The remarkable feature of Eq.~(\ref{lemotjuste}) which is not present in its 2DEG equivalent with parabolic dispersion is that the left hand side is simply
the average total energy, such that
\begin{equation}
\label{TheSolution}
\bar{E} = -r_s \frac{\partial \bar{E}}{\partial r_s}.
\end{equation}
Thus, the average total energy for many-body
systems with a linear dispersion interacting via a Coulombic potential satisfies a first
order differential equation; the solution, Eq.~(\ref{soln}), has many implications including those discussed
below~\footnote{Note that the result is exactly the same when using the Dirac equation for
ultra-relativistic fermions $\mathbf{H}_D=v_f \mathbf{\alpha}\cdot\mathbf{p} + V(\mathbf{r})$.}.  Eq.~(\ref{TheSolution})
is valid for all $d$, but other thermodynamic properties will have some explicit dependence on the dimension.

Following the work of Ceperley, as in~\cite{Ceperly}, it can easily be shown that
\begin{equation}
P \Omega=-\frac{r_s}{d}\frac{\partial E}{\partial r_s},
\end{equation}
where $P$ is the pressure, $\Omega$ is the generalized volume and $d$ is the dimension.  For a linear dispersion,
we have
\begin{equation}
P \Omega =\frac{T+V}{d}=\frac{E}{d},
\end{equation}
where again we have a simplification leading to a direct relationship involving the total energy $E$ and, in this case,
thermodynamic quantities.  This can be readilly used in a calculation of the chemical potential for graphene as well as other Dirac materials in arbitrary dimension via the Gibbs free energy
\begin{equation}
G = E + P \Omega = \left(1 + \frac{1}{d}\right)E = \mu N,
\end{equation}
so,
\begin{equation}
\mu = \left(1 + \frac{1}{d}\right)\frac{\mathcal{B}}{N r_s}
\end{equation}

From the expression for the total energy, Eq. (1), we can derive the electronic compressibility $\kappa$ 
 for $T = 0$, constant $N$ and $d = 1$, 2 and 3.  We will focus first on the compressibility for
 2D since this case, with a spin degeneracy of 2 and valley degeneracy of 2, models the Dirac material graphene.
 To compare the theoretical prediction of $\kappa$ with the measurements done on graphene [2], it is convenient
 to work with the energy per unit area, $\mathcal{E}=E/A$, we then find that 
\begin{eqnarray}
\label{IConeover}
\frac{1}{\tilde{\kappa}}&=&\frac{\partial^2\mathcal{E}}{\partial n^2} \nonumber \\
&=&\frac{\mathcal{B}a_0}{N}\frac{3}{4}\sqrt{\frac{\pi}{n}}
\end{eqnarray}
where we have introduced the expression $\tilde{\kappa}=n^2 \kappa$ and $n=N/A$.
Our result, Eq.~(\ref{IConeover}), which follows directly from the Virial Theorem, shows that
$\tilde{\kappa}^{-1} \propto n^{-1/2}$, as seen in experiment~\cite{Martin}, since
$\mathcal{B}$ is independent of $n$.  In their paper, Martin, \textit{et al.}~\cite{Martin} have shown 
that they can fit their data using the 2D free Fermi gas with a Dirac spectrum and a spin $+$ valley degeneracy of 4.
In this limit their only fitting parameter is the velocity, $v_g$, where $v_g \sim 10^6$ m/s is close to the value
obtained from the band structure~\cite{Wallace}.

While this interpretation is consistent with Eq.~(\ref{IConeover}) we need to look more closely at the interpretation of Eq.~(\ref{IConeover}) from
the point of view of the 2D Fermi liquid theory to get a better understanding of the role of interactions in the
Dirac materials.  The derivation of $\tilde{\kappa}$ in a 2D Fermi liquid with a Dirac spectrum is a straightforward
generalization of the 3D Landau Fermi liquid result~\cite{BaymPethick, Freedman},
\begin{equation}
\frac{1}{\tilde{\kappa}}=\frac{1+F_0^S}{N(0)},
\end{equation}
where the density of states is given by $N(0)=\frac{2 p_F}{\pi v^*_F}$ with $v^*_F$ the renormalized Fermi velocity
and $F_0^S$ is the spin symmetric quasiparticle interaction~\cite{BaymPethick}.  The Fermi momentum is related to the density by
$n = k_F^2/\pi$, thus $(1+F_0^S)v_F^* = (3B/2N)a_0$ must be independent of the density to be consistent with
Eq.~(\ref{IConeover}).  When comparing Eqs. (15) and (16) it is clear that only measuring $\tilde{\kappa}$
is not sufficient to determine the renormalizations of $v_F^*$ and $1+F_0^S$ independently.
 To determine the strength of the interaction parameter $F_0^S$ we need an independent measurement of the density
 of states. This can be obtained from the low temperature electronic specific heat at constant volume,
 \begin{equation}
 c_v=\frac{\pi^2}{3}N(0)T.
 \end{equation}
With a measurement of the specific heat we can answer the question: Do many body effects, coming from electron-electron
interactions play an important role in graphene?

In lieu of the measurements, we can make some comments regarding interaction effects in graphene.
In reference~\cite{HariWigner} Dahal \textit{et al.} argued that Wigner crystallization could not occur in graphene based on the Hartree-Fock
results for the kinetic $\bar{T}/N$ and potential $\bar{V}/N$ energy per particle, where
\begin{equation}
\label{TandV}
\bar{T}=\frac{2\sqrt{\pi}}{3}\hbar v_g n^{1/2};\qquad\bar{V}=\frac{4 \sqrt{\pi} e^2}{3 \epsilon} n^{1/2}.
\end{equation}
$e$ is the charge of the electron and $\epsilon$ is the dielectric of the medium calculated in the RPA~\cite{HariWigner}
approximation and is independent of $r_s$.  They found~\cite{HariWigner} that the ratio
$\bar{T}/|\bar{V}| \sim 1$ is independent of $r_s$.   Thus, the coefficient $\mathcal{B}$ calculated in the H-F approximation,
$\mathcal{B}^{\text{HF}} = \frac{2}{3}\frac{\hbar v_g}{a_0} - \frac{4}{3}\frac{e^2}{\epsilon a_0}$ is positive and
independent of $r_s$ but is much smaller than the experimental
value~\cite{Martin}.  To improve on the H-F approximation for $\mathcal{B}^{\text{HF}}$, one needs to include higher order
corrections to the ground state energy, for example, the RPA corrections used in reference~\cite{Barlas}.  Alternatively,
one could do the H-F approximation for the self-energy~\cite{Hwang}, which gives rise to a renormalization of the Fermi
velocity $v_F^*$.  While the approaches used in~\cite{Barlas} and~\cite{Hwang}, are clearly the right thing to do, the end result is,
both approaches give rise to a $\ln(1/r_s)$ contribution to the compressibility that is not seen experimentally.

The approximations, H-F and RPA, used in references~\cite{Barlas},~\cite{Hwang} and~\cite{HariWigner} for calculating the energy and the
compressibility have worked well as starting points in the traditional 2DEG and 3DEG.  The reasons for this
are: 1) The H-F approximation is considered to be exact in the low-density limit ($r_s \gg 1$)
and  2) The RPA is considered to be exact for the high-density limit ($r_s \ll 1$).  Thus, in the
traditional Coulomb gas, we have regimes of density where we can add perturbative corrections to these leading
order exact results.  From reference~\cite{HariWigner} and more precisely from Eq.~(\ref{soln}) it is clear that there is no perturbative
regime for graphene, or more generally for the case of a charged 2 dimensional Dirac gas (2DDG), since $\bar{T}/|\bar{V}|\sim 1$ for all $r_s.$
What we have learned from the virial theorem is that it is not surprising that the perturbation theory based efforts~\cite{Barlas, Hwang, HariWigner} to calculate the compressibility have not agreed with experiment.  Most likely calculating the
energy and/or the compressibility in a 2DDG would require a numerical calculation or a non-perturbative many-body approach,
along the lines of Ref.~\cite{Quader}. 

The effects of the interactions are even more striking when we consider the 1 dimensional Dirac gas (1DDG).
This is not just a theoretical toy model since it is possible to make a 1DDG by rolling up a graphene sheet
along specific directions, forming a metallic, 1D, single walled carbon nanotube~\cite{Kempa, Saito}.
There is a lot that is known about interacting Fermions in 1D, see for example
reviews in Refs.~\cite{Solyom, Deshpande}.  One of the more important results is that in 1D the interactions between
the fermions, no matter how weak, turn a Fermi gas into a Luttinger liquid, see for example Refs.~\cite{Tomonaga, Luttinger, Dzyalo, Solyom}.   Another important result is that in the case of the 1DEG there is no
Wigner crystal formation at any density~\cite{Schulz}.  This is expected since a long range
ordered state, like a Wigner crystal, cannot exist in 1D~\footnote{There are, however, density correlation
effects around a wave vector $q \sim 4k_F$ that decay very slowly, suggestive of Wigner crystal correlations~\cite{Schulz}
but this is not a true long range ordered state.}.

The absence of the Wigner crystal can be seen more clearly in terms of the energy density $\mathcal{E}=E/L$, where $E$ is given by Eq.~(\ref{soln}),
\begin{subequations}
\begin{eqnarray}
\mathcal{E}=\frac{\mathcal{B}}{N}a_0 n^2, \label{arraya}
\\
\frac{1}{\tilde{\kappa}}=2\frac{\mathcal{B}}{N}a_0. \label{arrayb}
\end{eqnarray}
\end{subequations}
We have defined $n = N/L$, with $L$ the length of the 1D chain, so $n=\frac{1}{r_s a_0}$ and $n=4k_F/\pi$.  Eq.~(\ref{arrayb}) for $\tilde{\kappa}$
is most interesting since it is independent of the density for 1D.  We expect this behavior to be seen in
single walled carbon nanotubes.  It turns out that there are models
in 1D that satisfy Eq.~(\ref{arrayb}), and one of them, we will see, is the Tomonaga-Luttinger model. In this
model there is an exact solution for the limit of extremely long-range interactions when the non-interacting
single particle spectrum is lineraized.  This problem can be solved by the standard bosonization techniques,
see, for example Ref.~\cite{Solyom}, turning the interacting fermions into non-interacting bosons

The boson velocity, i.e., the speed of 1st sound, $c_1$, is given by, see~\cite{Dzyalo},
$c_1^2 = v_F^2 + (2v_F/\pi) \lambda$, where $v_F=k_F/m$ and $m$ is the
bare electron mass.  Here the electron-electron interaction in momentum space is approximated by a coupling
constant $\lambda$ which is taken to be independent of the density and drops off rapidly for momentum transfers
$q \ll k_F$. If we plug this expression for $c_1^2$~\cite{Dzyalo} into the compressibility $\tilde{\kappa}^{-1}=(m/n)c_1^2$,
we get a $\tilde{\kappa}^{-1}$ that
depends on the density.  If we instead use the Dirac spectrum where $v_F$ is replaced by $v_g$ and $k_F/v_g$
replaces $m$, we get $\tilde{\kappa}^{-1}=(k_F/ v_g n)c_1^2$.  We find a density independent compressibility and
first sound, $c_1^2$;
we also find an energy density that is proportional to $n^2$~\footnote{The 1D result for the Tomonaga-Luttinger model with the Dirac spectrum suggests that our virial theorem,
while derived for the $1/r$ Coulomb potential also encompasses other examples of long-range forces.}.

From the Tomonaga-Luttinger model, there is an insight here that is relevant to the 2DDG.  If we calculate the energy density and
compressibility for the 1D free Dirac gas we would get that $\mathcal{E} \propto v_g n^2$ and
$\tilde{\kappa}^{-1}\propto v_g$, which has the same density
dependence as the interacting model.  Clearly, we cannot conclude that the interactions are not important simply
because we can get the same result as the Tomonaga-Luttinger model by simply rescaling the Fermi velocity in the
non-interacting Dirac gas.  The point here is that in the 1DDG, no matter how weak the interactions, the ground
state of this model is not a renormalized Fermi liquid, but in fact a Luttinger liquid.

Furter evidence of the power of the virial theorem in Dirac materials can be seen completely outside of
condensed matter, via studying weakly interacting quarks in the massless limit as in the MIT bag model~\cite{BagModel}.  Baym and Chin~\cite{BaymChin}
find that to order $g_c^2$ in the color coupling constant $g_c$, the speed of sound $c_1$ is independent of
density, precisely as predicted by the virial theorem for a linear dispersion in 3D.

In this letter we have presented the virial theorem for linear kinetic and Coulombic potential energies,
yielding a simple differential equation for the total energy, and we predict that $E\sim1/r_s$ for all densities
in Dirac Materials.  We have presented the wide applicability
of this powerful new solution in 1, 2 and 3 dimensions and it has been shown to agree with the experiment for the compressibility in 2D over a wide density range.  We have suggested that the results required by the virial theorem
for the linear spectrum with a $1/r$ potential apply to other potentials and presented systems in which
this is the case.

One of the authors (KSB) would like to thank Krzysztof Kempa for valuable discussions.  This work was supported, in part,
by the Center for Integrated Nanotechnologies, a U.S. Department of Energy, Office of Basic Energy Sciences
user facility.  This work was supported, in part, by a grant from Brookhaven National Laboratory.


\begin{thebibliography}{99}
  
  \bibitem{Wallace}
  P.~R.~Wallace, Phys. Rev. \textbf{71}, 622 (1947).
  
  \bibitem{Novo}
  K.~S.~Novoselov, \textit{et al.}, Science \textbf{306}, 666 (2004).  
  
  \bibitem{AIP}
See, for example, D.~S.~L.~Abergel, \textit{et al.}, Adv. Phys. \textbf{59}, 261 (2010); A.~K.~Geim and K.~S.~Novoselov, Nat. Mater. \textbf{6}, 183 (2007), and the refrences
contained within for a review.

  \bibitem{DiracMaterials}
  A.~Balatsky and T.~Wehling, unpublished. 
  
  \bibitem{Martin}
J.~Martin, \textit{et al.}, Nat. Phys. \textbf{4}, 144 (2008).

  \bibitem{Barlas}
Y.~Barlas, \textit{et al.}, Phys. Rev. Lett. \textbf{98}, 236601 (2007).

   \bibitem{Hwang}
E.~H.~Hwang, \textit{et al.}, 
Phys. Rev. Lett. \textbf{99}, 226801 (2007).

  \bibitem{HariWigner}
H.~P.~Dahal, \textit{et al.}, 
Phys. Rev. B \textbf{74}, 233405 (2006).

  \bibitem{Argyres}
P.~N.~Argyres, Phys. Rev. \textbf{154}, 410 (1967).

  \bibitem{March}
N.~H.~March, Phys. Rev. \textbf{110}, 604 (1958).

  \bibitem{Ceperly}
D.~Ceperley, Phys. Rev. B \textbf{18}, 3126 (1978).

  \bibitem{BaymPethick}
  G.~Baym and C.~Pethick, \textit{Landau Fermi Liquid Theory} (Wiley, New York 1991); D.~Pines and P.~Nozi\'eres, \textit{The Theory of Quantum Liquids} (Benjamin, New York, 1966), Vol. I.
  
  \bibitem{Freedman}
  R.~Freedman, Phys. Rev. B \textbf{18}, 2482 (1978).
  
  \bibitem{Quader}
  K.~F.~Quader, \textit{et al.}, 
Phys. Rev. B \textbf{36}, 156 (1987).

 \bibitem{Kempa}
 K.~Kempa, Phys. Rev. B \textbf{66}, 195406 (2002)
 
 \bibitem{Saito}
 R.~Saito, G.~Dresselhaus, and M.~S.~Dresselhaus, \textit{Physical Properties of Carbon Nanotubes}
 (Imperial College Press, London, 1999)
 
 \bibitem{Deshpande}
 V.~V.~Deshpande, \textit{et al}, Nature \textbf{464}, 209-216 (2010).
 
 \bibitem{Tomonaga}
 S.~Tomonaga, Prog. Theor. Phys. \textbf{5}, 544 (1950).
 
 \bibitem{Luttinger}
 J.~M.~Luttinger, J. Math. Phys. \textbf{4}, 1154 (1963).

  \bibitem{Dzyalo}
  Dzyaloshinskii and Larkin, Sov. Phys. JETP \textbf{38}, 202 (1974).
  
  \bibitem{Solyom}
  J.~Solyom, Adv. Phys \textbf{28}, 201 (1979).
  
      \bibitem{Schulz}
  H.~J.~Schulz,
Phys. Rev. Lett. \textbf{71}, 1864 (1993).


  \bibitem{BagModel}
  A.~Chodos, \textit{et al.}, Phys. Rev. D \textbf{9}, 3471 (1974).

 
  \bibitem{BaymChin}
  G.~Baym and S.~A.~Chin, Nucl. Phys. \textbf{A262}, 527-538 (1976).



\end{thebibliography}
\end{document}